\documentclass[journal=langd5,manuscript=article]{achemso}
\usepackage[version=3]{mhchem} 
\usepackage{natbib}
\usepackage{url}
\usepackage{appendix}

\usepackage{epstopdf}
\usepackage{mathtools}
\usepackage{textgreek}

\usepackage{multirow}
\usepackage{multicol}
\usepackage{graphicx}
\usepackage{caption}

\usepackage{amssymb}
\usepackage{float}
\usepackage[export]{adjustbox}
\usepackage{slashbox}

\usepackage{xcolor}

\newcommand{\ks}{\textcolor{black}} 

\author{Hari Govindha A.}
\affiliation{Department of Mechanical and Aerospace Engineering, Indian Institute of Technology Hyderabad, Kandi - 502284, Telangana, India}
\author{Pallavi Katre}
\affiliation{Department of Chemical Engineering, Indian Institute of Technology Hyderabad, Kandi - 502284, Telangana, India}
\author{Saravanan Balusamy}
\affiliation{Department of Mechanical and Aerospace Engineering, Indian Institute of Technology Hyderabad, Kandi - 502284, Telangana, India}
\email{saravananb@mae.iith.ac.in}
\author{Sayak Banerjee}
\affiliation{Department of Mechanical and Aerospace Engineering, Indian Institute of Technology Hyderabad, Kandi - 502284, Telangana, India}
\email{sayakb@mae.iith.ac.in}
\author{Kirti Chandra Sahu}
\affiliation{Department of Chemical Engineering, Indian Institute of Technology Hyderabad, Kandi - 502284, Telangana, India}
\email{ksahu@che.iith.ac.in}
\title{Counter-intuitive evaporation in nanofluids droplets due to stick-slip nature}

\begin{document}

\maketitle
\begin{abstract}
We experimentally investigate the evaporation characteristics of a sessile ethanol droplet containing Al$_2$O$_3$ and Cu nanoparticles of sizes 25 nm and 75 nm on a heated substrate using shadowgraphy and infrared imaging techniques. Our results demonstrate that the droplet contact line dynamics resulting from the presence of various nanoparticles plays a dominant role in the evaporation process. This is in contrast to the widely-held assumption that the enhanced evaporation rate observed in sessile nanofluid droplets is due to the higher thermal conductivity of the added nanoparticles. We observe that even though the thermal conductivity of Al$_2$O$_3$ is an order of magnitude lower than that of Cu, droplets containing 25 nm-sized Al$_2$O$_3$ exhibit pinned contact line dynamics and evaporate much more rapidly than droplets containing Cu nanoparticles of both sizes and 75 nm Al$_2$O$_3$ nanoparticles that exhibit stick-slip behaviour. We also found that the droplets with different nanoparticles display distinct thermal patterns due to the difference in contact line behaviour, which alters the heat transfer inside the droplets. We establish this counter-intuitive observation by analysing the temporal variations of the perimeter, free surface area, and deposition patterns on the substrate.
\end{abstract}


\section{Introduction}
\label{sec:intro}
Evaporation of sessile droplets laden with nanoparticles is relevant in a wide range of practical applications, such as inkjet printing \cite{de2004inkjet,tekin2004ink,soltman2008inkjet,park2006control}, fabrication of DNA microarrays \cite{dugas2005droplet,lee2006electrohydrodynamic}, estimating the lifetime of saliva droplets \cite{balusamy2021lifetime}, coating technology \cite{kim2016controlled,pahlavan2021evaporation}, spray and hotspot cooling \cite{kim2007spray,ruan2019effects,cheng2010active}, microfluidics \cite{deegan1997capillary}, to name a few. Additionally, this subject attracted the attention of researchers due to the profound scientific curiosity to understand the underlying mechanism of the resulting deposition patterns, including the commonly observed ``coffee-stain" or ``coffee-ring" effect \cite{deegan1997capillary,deegan2000contact,deegan2000pattern}. It is a prevalent belief that increasing the thermal conductivity of a liquid increases the heat transfer rate \cite{eapen2007mean,cui2022enhanced,zhang2021anisotropic}. Thus, the addition of nanoparticles in working liquids to enhance their thermal conductivity is a common strategy that has been employed for a long time in various applications \cite{choi1995enhancing,bi2008application,jang2006cooling}. 

Many researchers theoretically and experimentally studied the evaporation dynamics of sessile droplets in the presence of nanoparticles at ambient and elevated temperatures. \citet{orejon2011stick} investigated the three-phase contact line dynamics for pure water and ethanol on different substrates of varying hydrophobicities and showed that more hydrophobic surfaces favour the depinning of the contact line. They performed experiments with TiO$_2$ -water nanofluids and showed that the stick-slip behaviour depends on the nanoparticle concentration. \citet{moffat2009effect} reported the enhancement of the stick-slip behaviour with increasing nanoparticle concentration in TiO$_2$-ethanol nanofluid on a silicon wafer coated with Polytetrafluoroethylene (PTFE). \citet{yunker2011suppression} eliminated the coffee ring effect to obtain uniform deposition using ellipsoidal particles. These particles, with their attractive long-range forces, form structures near the contact line that prevents further deposition near the contact line. \citet{dugyala2014control} studied the effect of particle shape using colloidal ellipsoids and reported that the patterns do not depend on particle shape but are rather influenced by the interactions between particle and substrate. \citet{nguyen2013evaporation} generated inner coffee ring deposits with dendritic architectures using silica nanoparticles owing to the secondary pinning of the contact line when the forces acting on the particles are balanced. \citet{kovalchuk2014evaporation} experimentally investigated the effect of nanoparticle concentration on the evaporation dynamics and found an increase in the overall rate of diffusive evaporation with an increase in nanoparticle concentration. The type and concentration of nanoparticles can also significantly impact the evaporation rate \cite{moghiman2013influence}. \citet{vafaei2009effect} observed that the contact angle of the drop increases with increasing nanoparticle concentration and particle size. In pendant droplets laden with aluminium nanoparticles, it was found that the evaporation rate decreases with an increase in the particle concentration from 0 to 3 wt.\% \citet{gerken2014nanofluid}, while the surface tension is independent of the particle concentration \cite{tanvir2012surface}. \citet{jung2010forces} analysed the forces acting on the nanoparticles during the evaporation of a droplet on a hydrophilic substrate and found that the particles mostly experienced drag and surface tension forces. \citet{chen2010effects} used clay, silver, and iron oxide nanoparticles during the evaporation of a pendant water drop and found that the evaporation rate can be increased or decreased depending on the concentration and the type of nanoparticle. All these studies considered the evaporation dynamics of nanofluid droplets at room temperature.

A few researchers have performed molecular dynamics simulations to study droplet evaporation. The effect of electric fields and surface temperature on the evaporation of ionic droplets was investigated by \citet{chatterjee2021molecular}. They found a critical value of the electric field beyond which the hydration effect due to the ions was suppressed. \citet{caleman2007evaporation} obtained a relationship between the type of ion and water droplet evaporation. They also observed that the presence of the sodium and chlorine ions reduces the evaporation, while the hydrogen ions do not alter the evaporation. \citet{chen2020molecular} adopted molecular dynamic simulation to study the bubble nucleation on non-uniform wettability substrates. They observed that the nucleation position migrated towards the hydrophilic region with increased substrate temperature.

A few researchers have also investigated the evaporation dynamics of nanofluid droplets at elevated substrate temperatures \cite{sefiane2009nanofluids,brutin2011infrared,karapetsas2016evaporation,patil2016effects,zhong2017deposition}. \citet{patil2016effects} studied the effect of substrate temperature, colloidal particle concentration, and wettability on evaporation dynamics and deposition patterns in sessile droplets. They observed a ring-type deposition pattern in the inner region at elevated temperatures. \citet{zhong2017deposition} found that increasing the substrate temperature from 10$^\circ$C to 50$^\circ$C for a graphite nanofluid droplet on a silicon wafer changes the deposition pattern from a uniform disk profile to a dual ring structure. By varying the substrate temperature from 25$^\circ$C to 99$^\circ$C, \citet{parsa2015effect} obtained uniform, dual ring, and stick-slip deposition patterns in a sessile water droplet containing copper-oxide nanoparticles. The changes in the substrate temperatures affect the interplay between the capillary and Marangoni flows, which alters the deposition patterns. \citet{sefiane2009nanofluids} experimentally investigated the evaporation and wetting dynamics of a sessile ethanol droplet laden with aluminium nanoparticles on a PTFE substrate. They found that although the surface tension remains unaffected by the presence of nanoparticles, the contact angle increases due to the modification of the solid-liquid interfacial tension. \citet{brutin2011infrared} used an infrared (IR) camera to visualize the thermal patterns during the evaporation of sessile droplets of different semi-transparent liquids and observed that the surface instability depends on the fluid considered. \citet{zhong2017stable} reported that the increasing substrate temperature enhances the thermocapillary instabilities and the temperature heterogeneity of hydrothermal waves. Using IR thermography, \citet{sefiane2013thermal} showed that the hydrothermal waves extend across the entire droplet volume, and the thermal patterns affect the overall heat flux distribution inside the droplet.

As discussed above, apart from the various factors that affect the evaporation rates and deposition patterns, the thermal conductivity of the substrate and nanoparticles are important for evaporation since it expedites heat transfer \cite{bazargan2016effect}. In this context, \citet{sobac2012thermal} experimentally investigated the effect of temperature and thermal properties of the substrate on the evaporation of a pinned water droplet.  \citet{ristenpart2007influence} correlated the relative thermal conductivity of the substrate and liquid to the direction of the thermal-Marangoni flow, which alters the resulting deposition patterns. The higher evaporation rates were observed on substrates with high thermal conductivity \cite{lopes2013influence}. It was found that the thermal conductivity of the liquid increases with the addition of nanoparticles \cite{saterlie2011particle,garg2008enhanced,abdul2014thermal}. This enhancement in the thermal conductivity of the nanofluids has been attributed to the dispersion and Brownian motion of the nanoparticles. \citet{warrier2011effect} investigated the effect of the size of silver nanoparticles in ethylene glycol on the resultant thermal conductivity and found that the thermal conductivity of the nanofluid increases with an increase in particle size. \citet{beck2009effect} also observed similar behaviour for alumina nanoparticles. \citet{patel2010experimental} reported that the thermal conductivity of nanofluids with metallic nanoparticles is significantly higher than that with oxide nanoparticles. The theoretical studies \cite{choi1995enhancing,ren2005effective} also reveal that the presence of metallic nanoparticles enhances the thermal conductivity of the base fluids. 

As the abovementioned literature review suggests, adding nanoparticles increases the thermal conductivity of a base fluid, which in turn accelerates evaporation. However, the universality of this result has been questioned by some researchers \cite{eapen2007mean}. Thus, it is important to understand the mechanism underlying improved heat transfer in the presence of different nanoparticles. In the present work, we investigate the evaporation dynamics of sessile ethanol droplets with and without nanoparticle loading using shadowgraphy and infrared (IR) imaging techniques. Four different nanoparticles, Al$_2$O$_3$ (25 nm), Cu (25 nm), Al$_2$O$_3$ (75 nm) and Cu (75 nm) of various sizes with varying concentrations have been considered. The captured images are post-processed using the \textsc{Matlab}$^{\circledR}$ and a machine learning technique in the framework of a Convolutional Neural Network (CNN) based on the U-Net architecture. It is found that the lifetime of the droplets is not significantly affected by the increase in nanoparticle concentration. The droplet laden with Al$_2$O$_3$ (25 nm) nanoparticles shows pinned behaviour, whereas droplets laden with Cu (25 nm), Al$_2$O$_3$ (75 nm) and Cu (75 nm) show stick-slip behaviour. Our results reveal that a droplet containing Al$_2$O$_3$ (25 nm) nanoparticles evaporates significantly faster than droplets containing Cu nanoparticles (25 nm and 75 nm) and Al$_2$O$_3$ (75 nm) nanoparticles. This counter-intuitive behaviour is dedicated to the droplet contact line dynamics due to the presence of different nanoparticles. Additionally, the droplets with different nanoparticles exhibit distinct thermal patterns, altering the heat transfer inside the droplets. 

\section{Experimental Methodology}
\label{sec:expt}
\subsection{Experimental Setup}
 
\begin{figure}[h]
\centering
\includegraphics[width=0.9\textwidth]{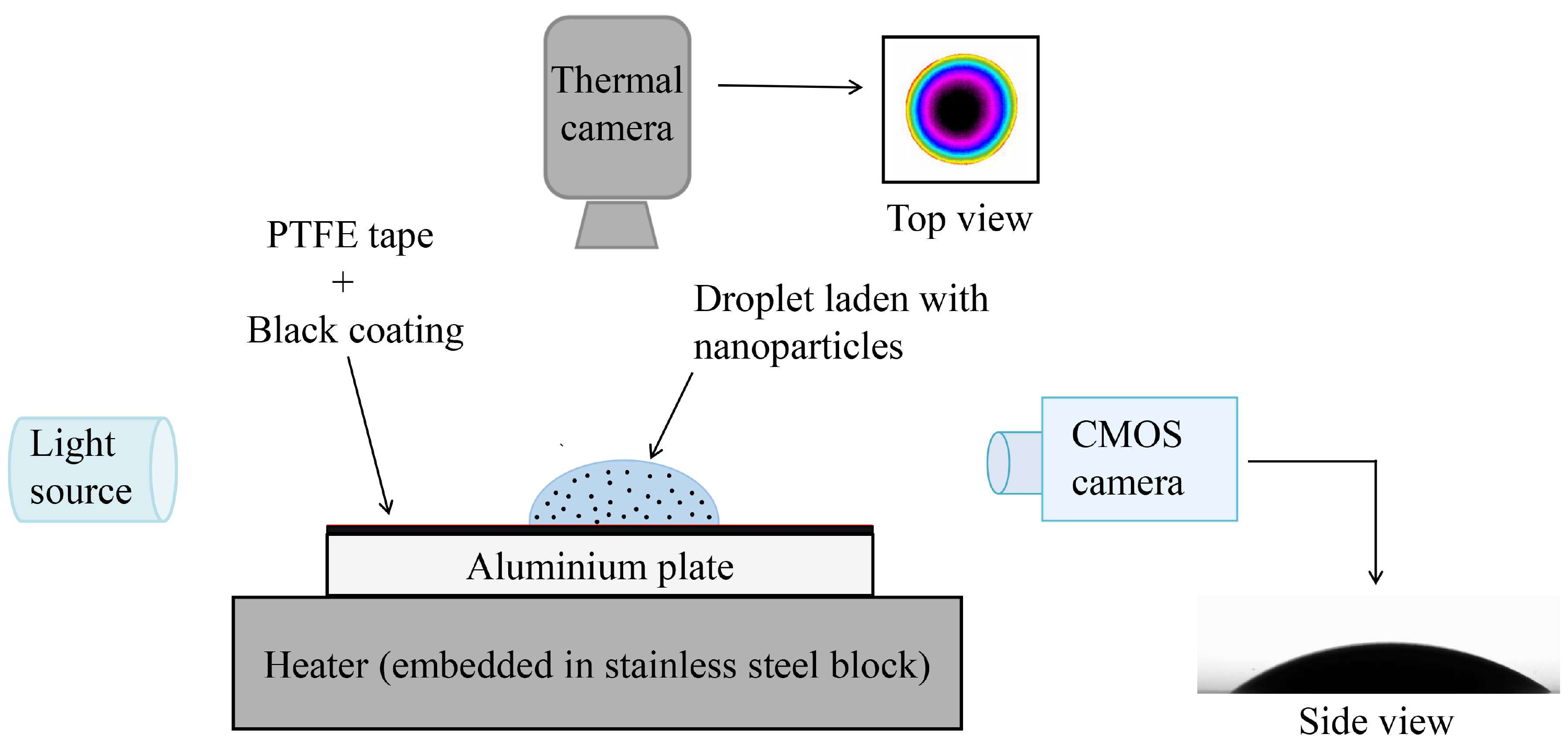}
\caption{Schematic diagram of the experimental setup (customized goniometer) to study the evaporation of sessile droplets laden
with nanoparticles.} 
\label{fig:fig1}
\end{figure}

We experimentally investigated the evaporation dynamics of a sessile ethanol droplet laden with different nanoparticles using shadowgraphy and infrared imaging techniques. The schematic diagram of the experimental setup is shown in Fig.\ref{fig:fig1}. The goniometer unit is customized for our requirements (Make: Holmarc Opto-Mechatronics Pvt. Ltd.). It consists of a multilayered metal block, a motor-driven pump for dispensing the droplets on the substrate, a proportional-integral-derivative (PID) controller for regulating the substrate temperature, a complementary-metal-oxide-semiconductor (CMOS) camera (Make: Do3Think, Model: DS-CBY501E-H), an infrared camera (Make: FLIR, Model: X6540sc), and an LED light source with a diffuser to distribute the light to the CMOS camera uniformly. The side and top views of the evaporating droplet were captured with the help of the CMOS and IR cameras, respectively. The entire assembly was placed inside the goniometer box to minimize external environmental disturbances. The goniometer box was maintained at an ambient temperature of $22\pm2^\circ$C and relative humidity of $45\pm5$\%. The relative humidity was measured using a hygrometer (Make: HTC, Model: 288-ATH) fitted inside the goniometer box.

The multilayered metal block consists of (i) a stainless steel base fitted with two electrical heaters operated by the Proportional–Integral–Derivative (PID) controller and (ii) an aluminum plate of size 100 mm $\times$ 80 mm $\times$ 15 mm coated with black paint to minimize the reflection in the IR images. A CMOS camera with a spatial resolution of $1280 \times 960$ pixels recorded the side view of the droplet at 10 frames per second (fps), which were used to extract various droplet parameters, such as the wetted diameter ($D$), height ($h$), contact angle ($\theta$) and volume ($V$). The IR camera captured the temperature distribution on the droplet surface from the top view with a resolution of $640 \times 512$ pixels at 50 fps in the spectral range of $3$ $\mu$m – $5$ $\mu$m. A polytetrafluoroethylene (PTFE) tape of thickness 100 $\mu$m is pasted on the aluminum plate, which is used as the substrate. The roughness and thermal stability of the PTFE tape were verified for the temperature range considered in this study \cite{katre2022experimental}. The required substrate temperature was obtained for each experiment by setting the PID controller and turning on the heater. A K-type thermocouple (Make: OMEGA Engineering Singapore) was used to check whether the substrate attained the steady state temperature before mounting the droplet. Before performing each experiment, it is ensured that the PTFE substrate is thoroughly cleaned with isopropanol, dried with compressed air, and then pasted onto the aluminum plate. The nanofluid solutions were prepared by dispersing the nanoparticles in absolute ethanol (99.9\% purity) on a weight percentage (wt.\%) basis. Then the mixture was ultrasonically shaken using an ultrasonic unit (Make: BRANSON, Model: CPX1800H-E) for about an hour, ensuring uniform nanoparticle distribution. Al$_2$O$_3$ and Cu nanoparticles with an average particle size of 25 nm and 75 nm were purchased from Sisco Research Laboratories Pvt. Ltd. and Intelligent Materials Pvt. Ltd., respectively. A 100 $\mu$L U-tek (Make: Unitek Scientific Corporation) chromatography syringe (with a piston size of 1.59 mm and fitted with a 21G needle with an inner orifice diameter of 0.514 mm) was connected to the motorised pump to control the volume flow rate, which in turn dispensed droplets of a constant size. A droplet of volume ($3.5 \pm 0.3$ $\mu$L) created using this mechanism was placed on the substrate, and its evaporation dynamics were recorded using the CMOS and IR cameras. In our experiments, time, $t=0$, is the instant when the droplet touches the substrate. After each experiment, the PTFE tape was replaced, and the syringe was cleaned with acetone. We have performed a minimum of three repetitions for each set of the experimental condition. A digital microscope (Make: Keyence, Model: VHX-6000) was used to examine the dried deposition pattern once evaporation was completed.

\subsection{Post-processing}
To extract the droplet side view profiles, the post-processing of the side view images recorded by the CMOS camera was performed using an in-house program in the framework of \textsc{Matlab}$^{\circledR}$. It was accomplished using a median filtering technique to eliminate random noise and an unsharp masking technique to sharpen the image, improving the gradients. The filtered image was then converted to a binary image using a suitable threshold that differentiates the background from the droplet boundary. Finally, the holes were filled inside the droplet boundary, and the reflection of the droplet was removed. A \textsc{Matlab}$^{\circledR}$ function was used to trace the droplet contour from which the droplet parameters were measured. \ks{Figure S1(a-e)} shows the steps followed in the image processing of the side-view images of the droplets. The detailed description of the post-processing procedure is similar to that of  \citet{gurrala2019evaporation}. In order to analyze infrared images, the intensity data of the image was converted to a temperature field \cite{katre2020evaporation}. The Convolution Neural Network based on U-net architecture was used for the boundary extraction\cite{katre2021evaporation}. The U-net design uses data augmentation by elastically deforming the annotated input photos, enabling the network to use the available annotated images more. The network was trained using 40 manually annotated grey-scale infrared images. A computer equipped with a GPU (NVIDIA Quadro P1000) was used for the training of the images. The network was then used to extract the binary masks and droplet boundaries from the infrared images, as shown in \ks{Figure S2}. Finally, a \textsc{Matlab}$^{\circledR}$ code was used to remove the background, and the temperature profiles of the evaporating droplets at different instants were analysed.

\section{Results and Discussion}\label{sec:results}
\subsection{Droplet lifetime ($t_e$)}
We investigate the evaporation dynamics of sessile ethanol droplets on heated substrates with and without Al$_2$O$_3$ and Cu nanoparticle loadings. The substrate temperature is kept at $T_s=50^\circ$C. The particle size and the particle loading concentration are varied, and the impact of particle type, particle size and concentration on droplet evaporation dynamics have been investigated. We consider two different mean diameter sizes for both the Al$_2$O$_3$, and Cu nanoparticles, viz. 25 nm and 75 nm. Four different particle loading concentrations, 0 wt.\%, 0.3 wt.\%, 0.6 wt.\% and 0.9 wt.\% are considered for each particle types and diameters. Table \ref{table:T1} shows the lifetime of the droplets for all the loading cases.
 
\begin{table}
\centering
\caption{Lifetime of an ethanol droplet (in seconds) laden with Al$_2$O$_3$ and Cu nanoparticles of different sizes and concentrations at $T_s=50^\circ$C.} 
\label{table:T1}
\hspace{2 mm}\\
\begin{tabular}{|c|c|c|c|c|c|c|}\hline
\multirow{2}{*}{\backslashbox{Size}{Concentration}}&\multicolumn{2}{|c|}{0.3 wt.\%} & \multicolumn{2}{|c|}{0.6 wt.\%} &\multicolumn{2}{c|}{0.9 wt.\%}\\
\cline{2-7}	
 & Al$_2$O$_3$ & Cu & Al$_2$O$_3$ & Cu & Al$_2$O$_3$ & Cu\\ 
\hline
25 nm & 43$\pm1$  & 65$\pm2$  & 43$\pm1$  & 64$\pm2$  & 44$\pm1$  & 63$\pm3$ \\
\hline
75 nm & 66$\pm2$  & 60$\pm3$  & 69$\pm4$  & 62$\pm1$  & 64$\pm6$  & 63$\pm3$  \\
\hline
\end{tabular}
\end{table}

We observe that the lifetime of a pure ethanol droplet is $74 \pm 3$ seconds. A comparison of the pure ethanol droplet lifetime with the lifetimes of the nanoparticle-laden droplets given in Table \ref{table:T1} reveals that the lifetime of the droplets is reduced by the addition of nanoparticles irrespective of particle type, particle diameter or extent of loading wt.\%. However, the extent of reduction in droplet lifetime varies significantly for the different cases. For Cu nanoparticles of both 25 nm and 75 nm mean diameters, the decrease in the lifetime is modest, with the total lifetimes varying between 87\% to 81\% of the pure droplet lifetime. The impact of increasing the particle concentration is also relatively small. The same statement holds for Al$_2$O$_3$ laden droplets where the mean particle size is at 75 nm, with the droplet lifetimes being about 86\% to 92\% of the pure droplet lifetime with no significant impact observed for increased particle concentrations. However, the state of affairs is markedly different when 25 nm sized Al$_2$O$_3$ nanoparticle-laden droplets are considered. For these cases and all concentrations, the droplet lifetime shows a marked decrease, reducing to 58\% of the lifetime of the pure ethanol droplet. Thus the Al$_2$O$_3$ laden nanoparticle droplets show a significant impact of particle size on the droplet evaporation time, which is not seen in the Cu nanoparticle case. Remarkably, the 25 nm sized Al$_2$O$_3$ nanoparticle case shows anomalously faster evaporation rates than all the other conditions. This finding appears counter-intuitive since the thermal conductivity of Cu nanoparticles are more than 10 times higher than that of Al$_2$O$_3$ nanoparticles (Table \ref{table:TS1}). Further investigations presented in this manuscript attempt to elucidate the reasons for this behavior. Since different particle loadings do not show any significant impact, the results discussed in the subsequent sections of this work deal with 0.6 wt.\% nanoparticle loading cases only. 

\begin{table}
\caption{The properties of nanoparticles at 27$^\circ$C \cite{PerryHandbook}.}
\hspace{0.0cm}   \\
\label{table:TS1}
\begin{tabular}{|c|c|c|c|c|c|}\hline
\multirow{2}{*}{Nanoparticle}  & Density & Thermal conductivity & Molar mass & Specific heat \\
 & (kg/m$^3$) & (W/mK) & (kg/kmol) & (J/kgK)\\\cline{1-5}
Al$_2$O$_3$ & 3.9 & 36 & 101.96 & 765 \\ \cline{1-5}
Cu & 8.9 & 401 & 63.55 & 385 \\ 
\hline
\end{tabular}
\end{table}

\subsection{Evaporation dynamics: Side view profiles}
\begin{figure}[h]
\centering
\hspace{0.5cm}{\large (a)}\\
\hspace{-0.2cm}\includegraphics[width=0.45\textwidth]{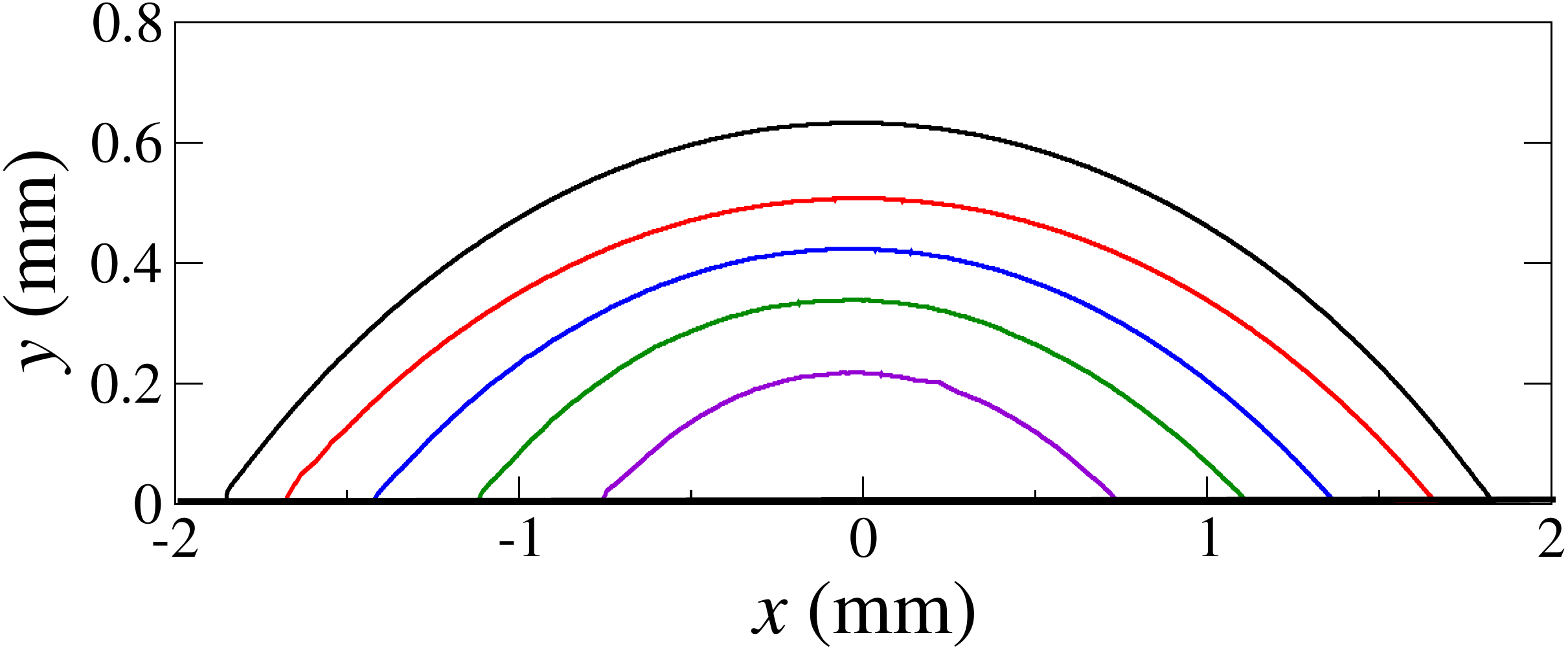} \hspace{2mm}\\
\hspace{0.5cm}{\large (b)} \hspace{7.1cm}{\large (c)} \\
 \includegraphics[width=0.45\textwidth]{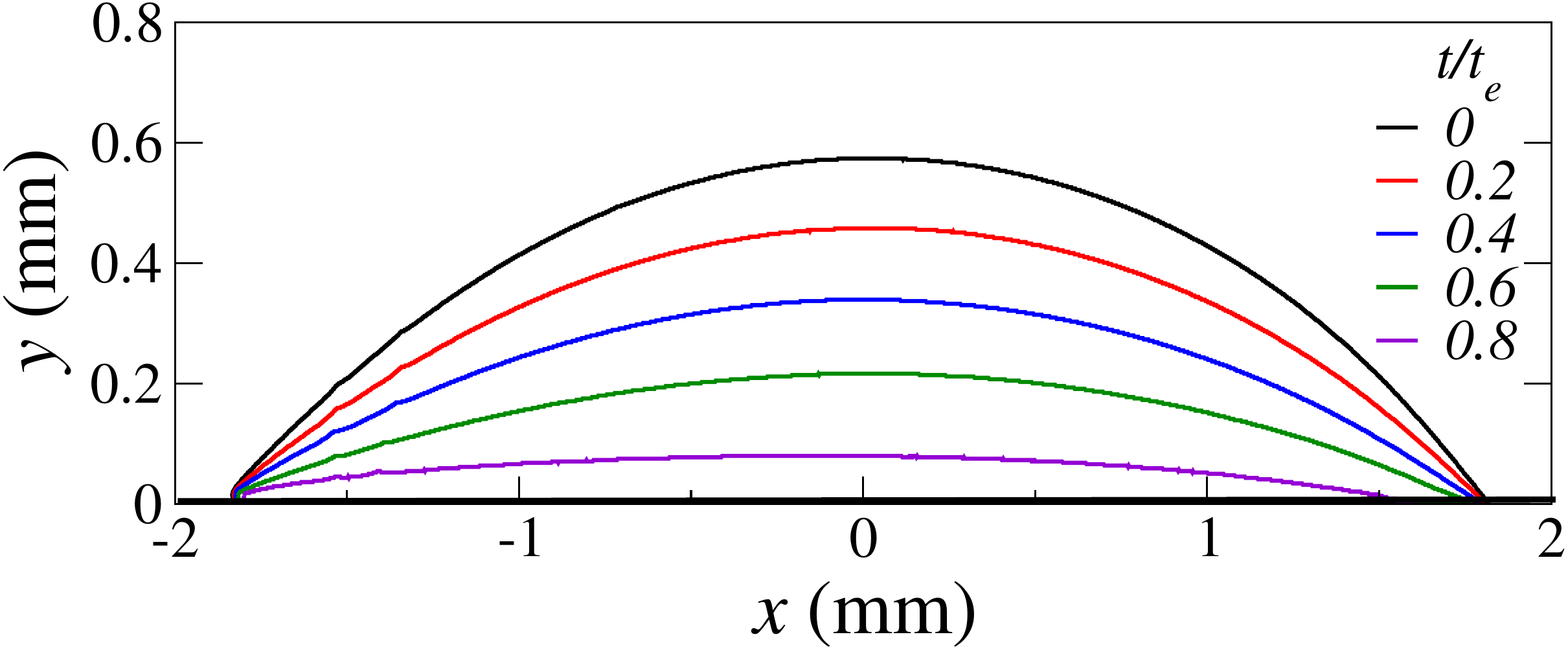} \hspace{2mm} \includegraphics[width=0.45\textwidth]{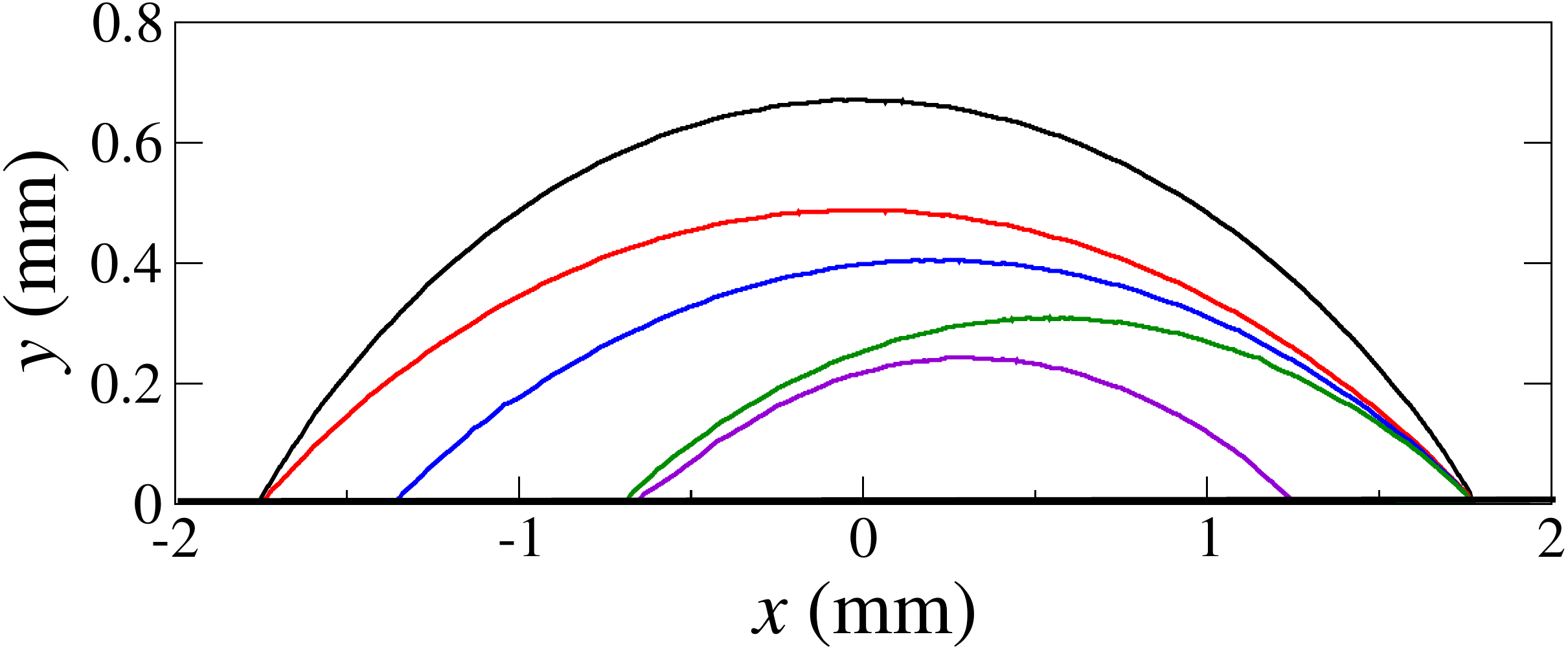}\\
 \hspace{0.5cm}{\large (d)} \hspace{7.1cm}{\large (e)} \\
 \includegraphics[width=0.45\textwidth]{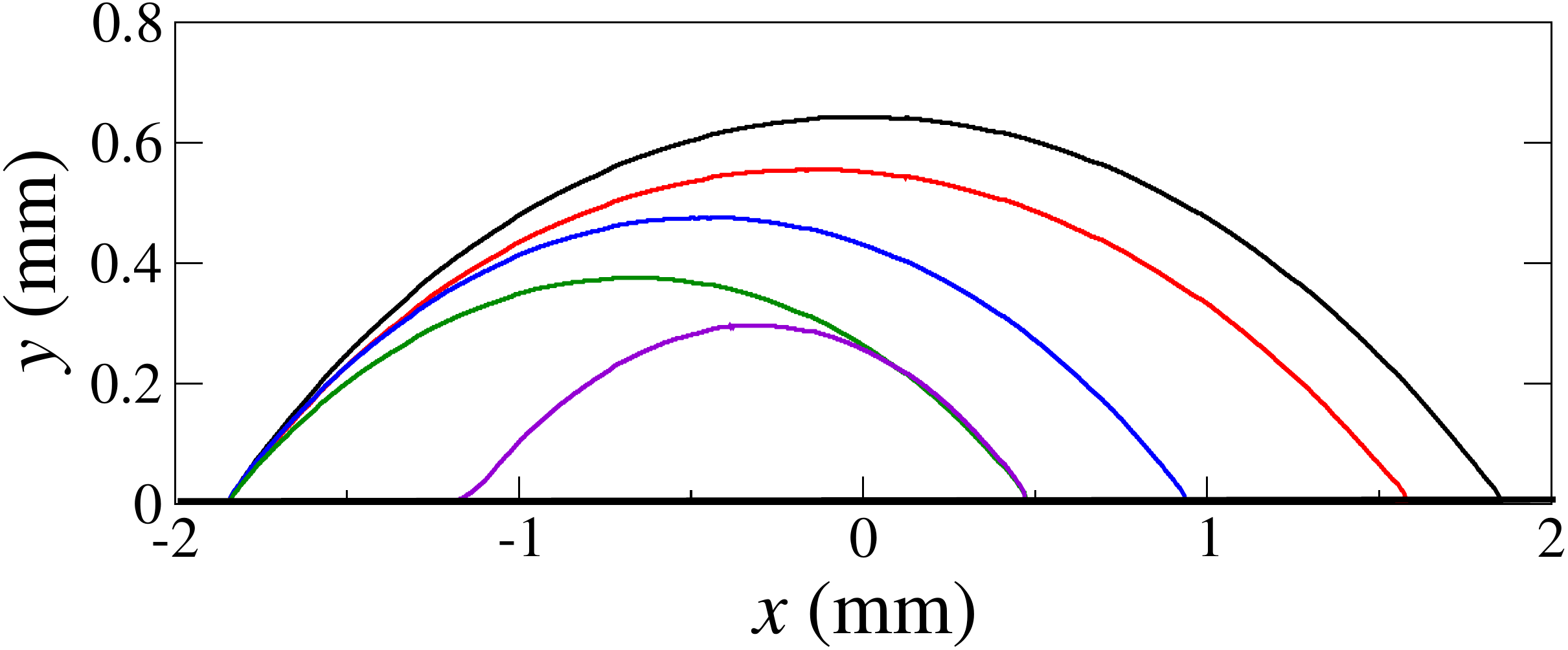} \hspace{2mm} \includegraphics[width=0.45\textwidth]{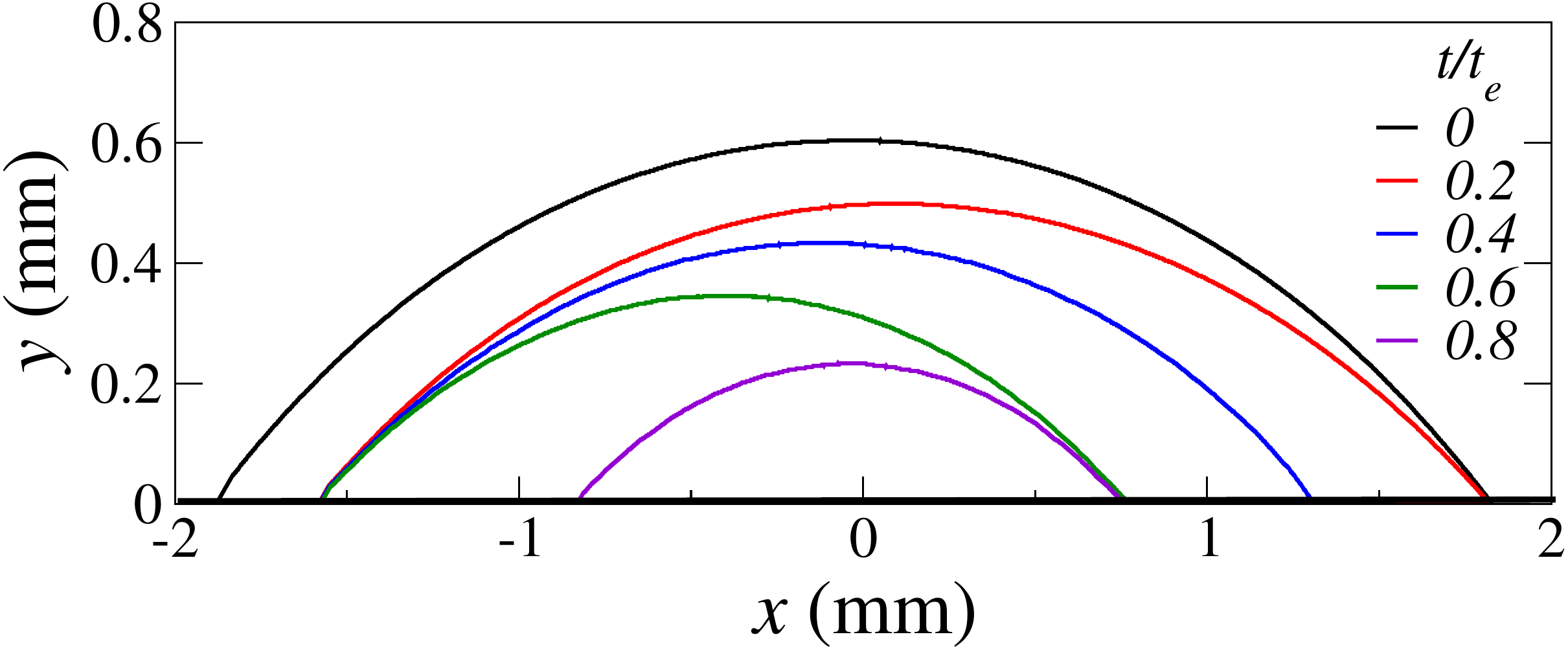} 
\caption{Temporal evolution of ethanol droplet contours (a) without nanoparticle loading, (b) 0.6 wt.\% loading of 25 nm Al$_2$O$_3$, (c) 0.6 wt.\% loading of 25 nm Cu, (d) 0.6 wt.\% loading of 75 nm Al$_2$O$_3$ and (e) 0.6 wt.\% loading of 75 nm Cu nanoparticles at $T_s=50^\circ$C.}
\label{fig:fig2}
\end{figure}

This section presents the temporal evolution of the side contour profile of an ethanol droplet with and without nanoparticles at 0.6 wt.\% loading. The experimental section has already discussed details of extracting contour profiles from CMOS camera data. The side-view droplet images for the various cases are shown in \ks{Figure S3}. Figure \ref{fig:fig2} shows the superimposed droplet contour profiles at different dimensionless time $(t/t_e)$, wherein $t_e$ is the lifetime for the given case. The $x$ axis provides a measure of the droplet spread, while the $y$ axis provides a measure of droplet height. The contours are provided from the initial time ($t/t_e = 0$) to 80\% of the droplet lifetime ($t/t_e = 0.8$). Figure \ref{fig:fig2}a depicts the side contour profiles of a pure ethanol droplet.  It can be seen that as $t/t_e$ increases, the droplet wetting diameter decreases monotonically and in a symmetrical fashion for an ethanol droplet without nanoparticle loading. This observation is consistent with the Constant Contact Angle (CCA) mode of evaporation, where a sessile droplet maintains a constant contact angle with respect to the substrate throughout its evaporation lifetime, which leads to a monotonic decrease in its wetting diameter.  
 
Figure \ref{fig:fig2}b gives the side contour profile for the droplets laden with 25 nm sized Al$_2$O$_3$ nanoparticles with 0.6 wt.\% loading. It can be observed that the evolution of the droplet side profile is dramatically different, with the droplet spread remaining more or less constant up to 80\% of the droplet lifetime and the droplet contact angle decreasing monotonically with time. The droplet evaporation behavior is therefore consistent with the Constant Contact Radius (CCR) mode of evaporation, where the droplet spread diameter remains constant throughout its lifetime. Hence, it can be concluded that the presence of 25 nm sized Al$_2$O$_3$ nanoparticles has resulted in a droplet pinning effect where the contact line cannot retract from its initial position despite progressive evaporation.
  
Figures \ref{fig:fig2}(c-e) show the droplet side profile evolutions for 25 nm Cu, 75 nm Al$_2$O$_3$ and 75 nm Cu loaded nanoparticle cases respectively, all at 0.6 wt.\% loading and at substrate temperature of $50^\circ$C. In all these cases, the contour evolution is irregular and asymmetric. The spread diameter and the contact angle show an irregular decrease with time. The rate of decrease is also different between the left and right sides, and thus, at several time points, the center of the droplet shifts leftward or rightward, depending on which edge has contracted the most. Such behaviour is characteristic of the stick-slip mode of droplet evaporation\cite{maheshwari2008coupling}.   
 
\begin{figure}[h]
\centering
\includegraphics[width=1.0\textwidth]{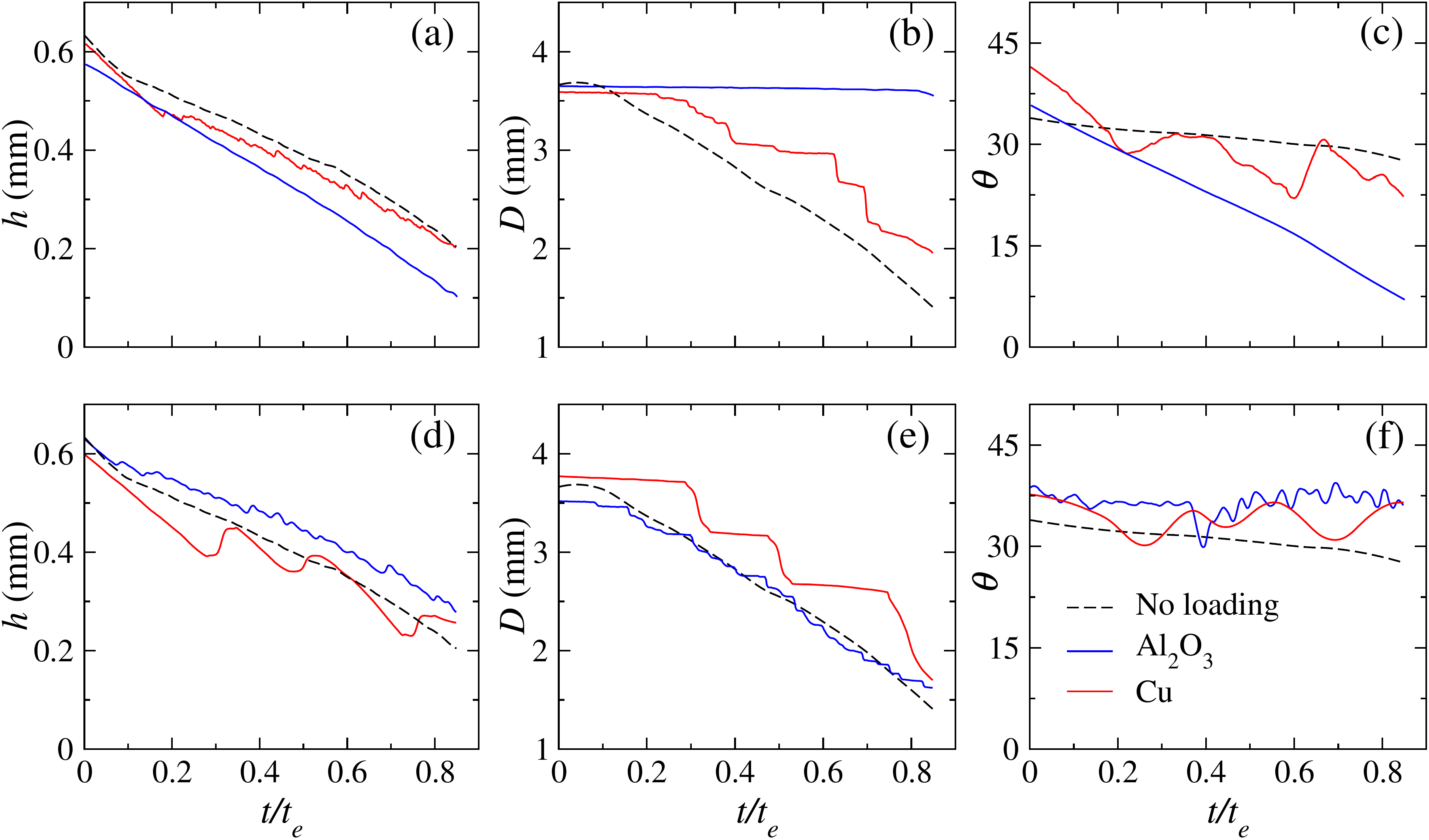}
\caption{Variations of the height ($h$ in mm), wetted diameter ($D$ in mm) and contact angle ($\theta$ in degree) of ethanol droplets at $T_s=50^\circ$C. The first row (a, b, c) and second row (d, e, f) represent the droplets containing nanoparticles of size 25 nm and 75 nm, respectively.}
\label{fig:fig3}
\end{figure}
 
The variations in the height ($h$ in mm), wetted diameter ($D$ in mm), and contact angle ($\theta$ in degree) of the droplet with and without nanoparticle loading are plotted with respect to normalized evaporation time in Figure \ref{fig:fig3}. The first row of figures, Fig. \ref{fig:fig3}(a-c), compares the no-loading condition with droplets having 25 nm Al$_2$O$_3$ and Cu nanoparticle loading cases. The second row of figures, Fig. \ref{fig:fig3}(d-f), compares the no-loading condition with droplets having 75 nm Al$_2$O$_3$ and Cu nanoparticle loading cases. It is seen that the wetted diameter of the pure ethanol droplet decreases monotonically while the contact angle remains constant. This is consistent with the CCA mode of evaporation, as noted earlier. For droplets with 25 nm Al$_2$O$_3$ loading, the wetted diameter is observed to be constant while the contact angle is observed to decrease monotonically, as expected for the CCR mode of evaporation. In contrast, the Cu (25 nm) droplet includes regimes of pinning, i.e. CCR mode evaporation, where the droplet diameter stays constant, and the droplet contact line decreases monotonically, that are followed by regions of de-pinning and droplet contraction, where the diameter of the droplet shows a steep decrease in a very short period and the contact angle shows a rapid increase. The de-pinning process ends within a short period, and the droplet stabilizes into a new pinned evaporation phase with a lower wetted diameter. The duration of the pinned phase is highly uneven, with some pinned phases lasting over significant fractions of the droplet lifetime followed by a de-pinning phase with abrupt and marked contraction `jump'. At other instances, the pinned-depinned regimes occur in rapid succession in micro-steps or `jerks' that almost replicate the CCA mode of droplet diameter evolution. Overall the 25 nm Cu nanoparticle-laden droplet exhibits a stick-slip mode.

Now, considering the droplet containing Al$_2$O$_3$ (75) nm and Cu (75 nm) nanoparticles in Figure \ref{fig:fig3}(d-f), it can be seen that both these droplets exhibit stick-slip behaviour. For the cases shown, the Al$_2$O$_3$ laden droplet evaporates through a series of micro-pinning and de-pinning processes that follow each other in rapid succession. Thus, its diameter and height evolution are close to the CCA mode exhibited by the pure droplet. In contrast, the Cu nanoparticle droplet passes through a few large and stable pinned phases, followed by rapid large contractions in droplet diameter accompanied by increases in droplet height and droplet contact angle. It is to be noted, however, that the stick-slip behavior can shift from micro pinning and de-pinning mode to long-duration pinned regimes for different runs of the same case, as is shown in \ks{Figure S4}. The number of stick slips and the time at which each stick-slip occurs are not the same for different individual runs of the droplets containing Cu (25 nm), Al$_2$O$_3$ (75 nm), and Cu (75 nm) nanoparticles. The variations in the wetted diameter of Al$_2$O$_3$ and Cu of different sizes are compared in \ks{Figure S5}, which clearly shows that the 25 nm Al$_2$O$_3$ nanoparticle case has the distinct pinned mode evaporation, whereas the rest have stick-slip mode evaporation. The thermal patterns and instabilities seen during droplet evaporation are analyzed in the next section. 

\subsection{Evaporation dynamics: Top view profiles}
\begin{figure}[h]
\centering
\includegraphics[width=1\textwidth]{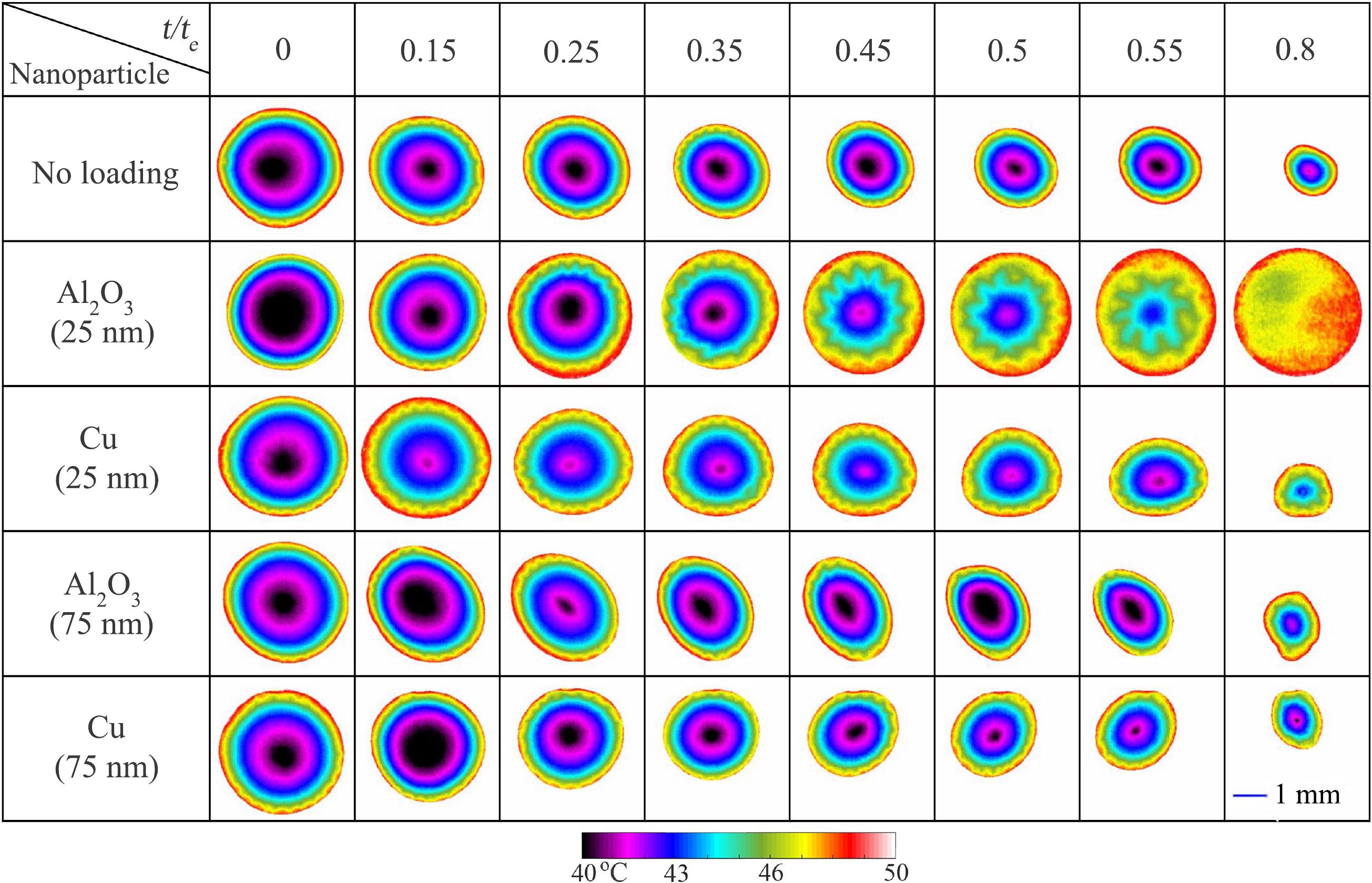} 
\caption{Temporal evolution of the temperature contours on the surface of the droplet in the no loading and 0.6 wt.\% loading conditions at $T_s=50^\circ$C. The compositions of droplet containing different nanoparticles are presented as: First row (no loading), second row (Al$_2$O$_3$ of 25nm), third row (Cu of 25nm), fourth row (Al$_2$O$_3$ of 75 nm) and fifth row (Cu of 75nm). The videos showing thermal profiles of the droplet for no loading, Al$_2$O$_3$ (25nm), Cu (25nm), Al$_2$O$_3$ (75nm) and Cu (75nm) are included as \ks{Videos S1-S5}, respectively.}
\label{fig:fig4}
\end{figure}

In this section, the temperature distribution on the droplet surface with and without nanoparticle loading is investigated using the IR camera mounted above the droplet in a plane perpendicular to the substrate. The top view of the ethanol droplets with and without nanoparticle loading is shown in Figure \ref{fig:fig4}.
 
In Figure \ref{fig:fig4}, it is evident that the droplet without loading shows a continuous decrease in the wetted diameter while the droplet with Al$_2$O$_3$ (25 nm) shows pinned behaviour. The other droplets with Cu (25 nm), Al$_2$O$_3$ (75 nm), and Cu (75 nm) display stick-slip behaviour and because of uneven pinning, deviate from a spherical cap profile in the later stages of their lifetimes. The hydrothermal waves, which originate from the droplet periphery, are observed in all the droplets as quasi-regular undulations in the iso-temperature profiles as viewed from the top. For the droplet with Al$_2$O$_3$ (25 nm) nanoparticles, the height decreases at constant wetting diameter due to pinning, which results in the propagation of hydrothermal waves towards the centre of the droplet. This promotes mixing inside the droplet, enhancing heat transfer and thereby decreasing the lifetime. This can be visualized by observing the central region of the droplet, which clearly shows a higher temperature compared to the central regions of the other droplets. Now, comparing the temperature profiles of droplets with Cu (25 nm), Al$_2$O$_3$ (75 nm), and Cu (75 nm), we can observe that the central regions of droplets with Cu (25 nm) and (75 nm) have a higher temperature than that of Al$_2$O$_3$ (75 nm). This slight increase could be attributed to the increased thermal conductivity of the Cu nanoparticles, which still could not significantly alter the lifetime of these droplets. From this, it can be realised that the evaporation dynamics are majorly affected by the contact line dynamics, due to which the pinned Al$_2$O$_3$ (25 nm) droplet has a lower lifetime, even though it has a lower conductivity compared to that of copper. A few repetitions of the thermal profiles showing stick-slip in different directions for the ethanol droplet with Cu (25 nm) and Cu (75 nm) are shown in \ks{Figures S6 and S7}, respectively.

\subsection{Perimeter and free surface area}
\begin{figure}[h]
\centering
\includegraphics[width=1\textwidth]{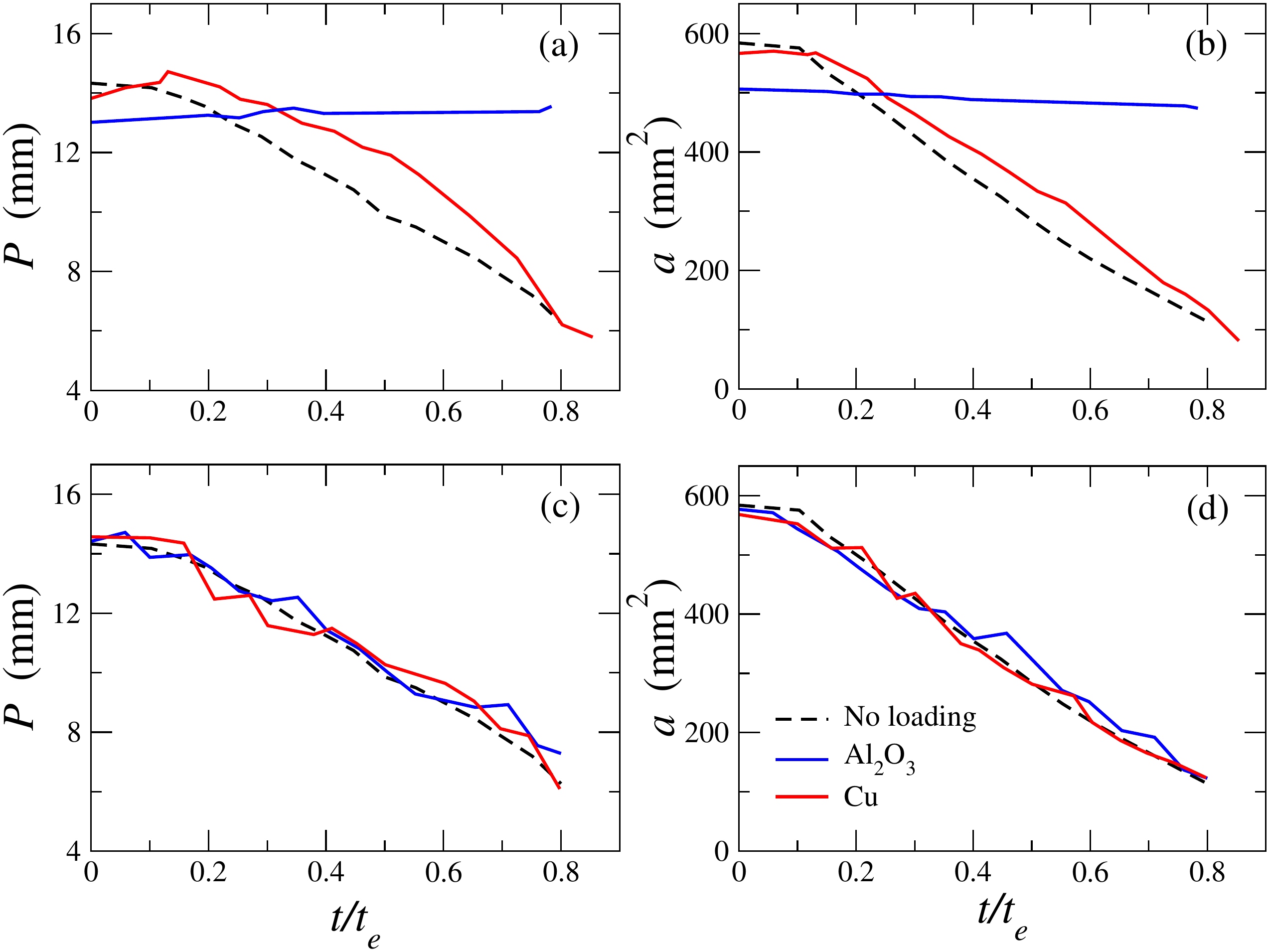}
\caption{Temporal variation of the perimeter ($P$ in mm) and surface area ($a$ in mm$^2$) of the droplets at $T_s=50^\circ$C. The first row (panels a, b) and second row (panels c, d) are associated with the droplets laden with nanoparticles of size 25 nm and 75 nm, respectively.}
\label{fig:fig5}
\end{figure}

Additional insight into the evaporation dynamics of the nanofluid droplets is obtained by plotting the temporal variations of its contact line perimeter and free surface area in the presence and absence of nanoparticle loading at $T_s=50^\circ$C. Figure \ref{fig:fig5}(a-b) and \ref{fig:fig5}(c-d) show the variations in the perimeter and free surface area of the pure ethanol droplet and droplets containing nanoparticles of size 25 nm and 75 nm, respectively. Many studies have shown that the evaporation of droplets primarily occurs at the triple contact line \cite{starov2009evaporation}. Moreover, for a heated droplet, due to natural convection, the evaporation flux depends on the surface area of the droplet \cite{gurrala2019evaporation}. The pure ethanol droplet exhibits the CCA mode of evaporation; hence, its perimeter and surface area decrease nearly monotonically with time. Hence the droplet evaporation rates also decrease along with the decrease in the droplet surface area and the triple contact line perimeter. In contrast, the droplet with 25 nm Al$_2$O$_3$ nanoparticles shows a pinning effect as they evaporate in the CCR mode, and hence the perimeter and the free surface area remain unchanged throughout its lifetime. Thus, during its lifetime, the Al$_2$O$_3$ (25 nm) droplets have a higher contact line perimeter and free surface area than all other droplets. Hence it is expected to have the highest evaporation rates and the shortest droplet lifetime. Due to the stick-slip nature of Cu (25 nm), Cu (75 nm) and Al$_2$O$_3$ (75 nm)  droplets, the perimeter and free surface area decline at rates slightly lower than the pure ethanol case, and hence they have a slightly higher evaporation rate compared to that of a pure ethanol droplet. These observations explain why the  Al$_2$O$_3$ (25 nm) laden droplets have the smallest droplet lifetimes, the pure ethanol droplets have the largest lifetimes, and the  Al$_2$O$_3$ (75 nm) and 25 nm and 75 nm Cu nanoparticle droplets have lifetimes slightly shorter than the pure ethanol case as was shown in Table \ref{table:T1}. The preceding discussion clarifies that, at least for low loading concentrations (up to 0.9 wt.\% nanoparticles), the thermal conductivity of nanoparticles does not play a significant role in determining the heat transfer and evaporation rates of sessile droplets from a heated substrate. Instead, the droplet evaporation rates and lifetimes are determined by how the nanoparticle loading affects the contact line dynamics of the evaporating droplet. 

\subsection{Deposition pattern and roughness profile}
\begin{figure}
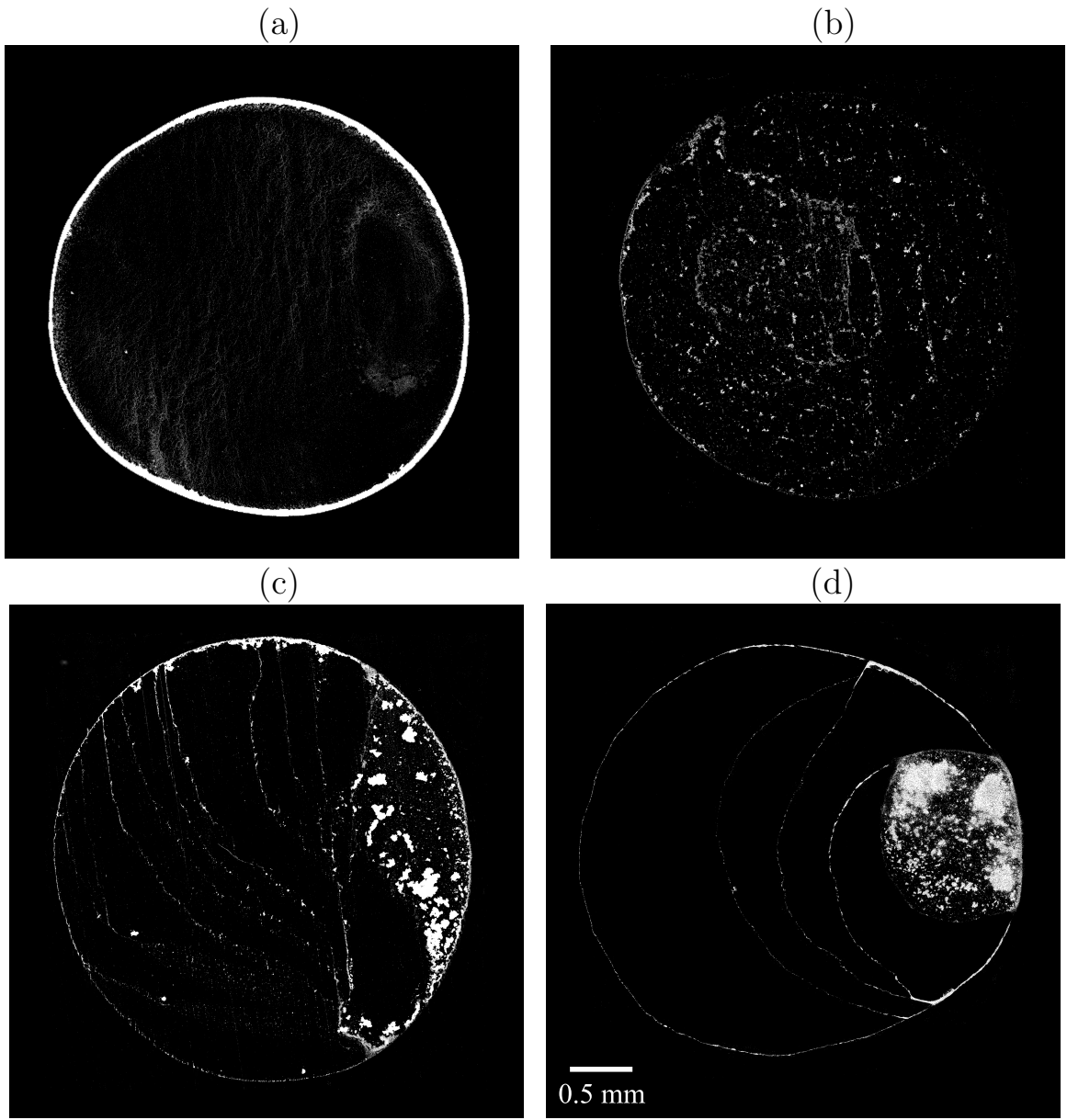

\centering
 \hspace{0.5cm}  {\large (a)} \hspace{7.1cm} {\large (b)} \\
 \includegraphics[width=0.45\textwidth]{Figure6a.pdf} \hspace{2mm} \includegraphics[width=0.45\textwidth]{Figure6b.pdf}\\
 \hspace{0.5cm}  {\large (c)} \hspace{7.1cm} {\large (d)}\\
 \includegraphics[width=0.45\textwidth]{Figure6c.pdf} \hspace{2mm}\includegraphics[width=0.45\textwidth]{Figure6d.pdf}
\caption{Deposition pattern of ethanol droplets laden with nanoparticles of Al$_2$O$_3$ (a,c) and Cu (b,d) at $T_s=50^\circ$C. Panels (a,b) and (c,d) correspond to the nanoparticles size of 25 nm and 75 nm, respectively.}
\label{fig:fig6}
\end{figure}

\begin{figure}
\centering
\includegraphics[width=1.0\textwidth]{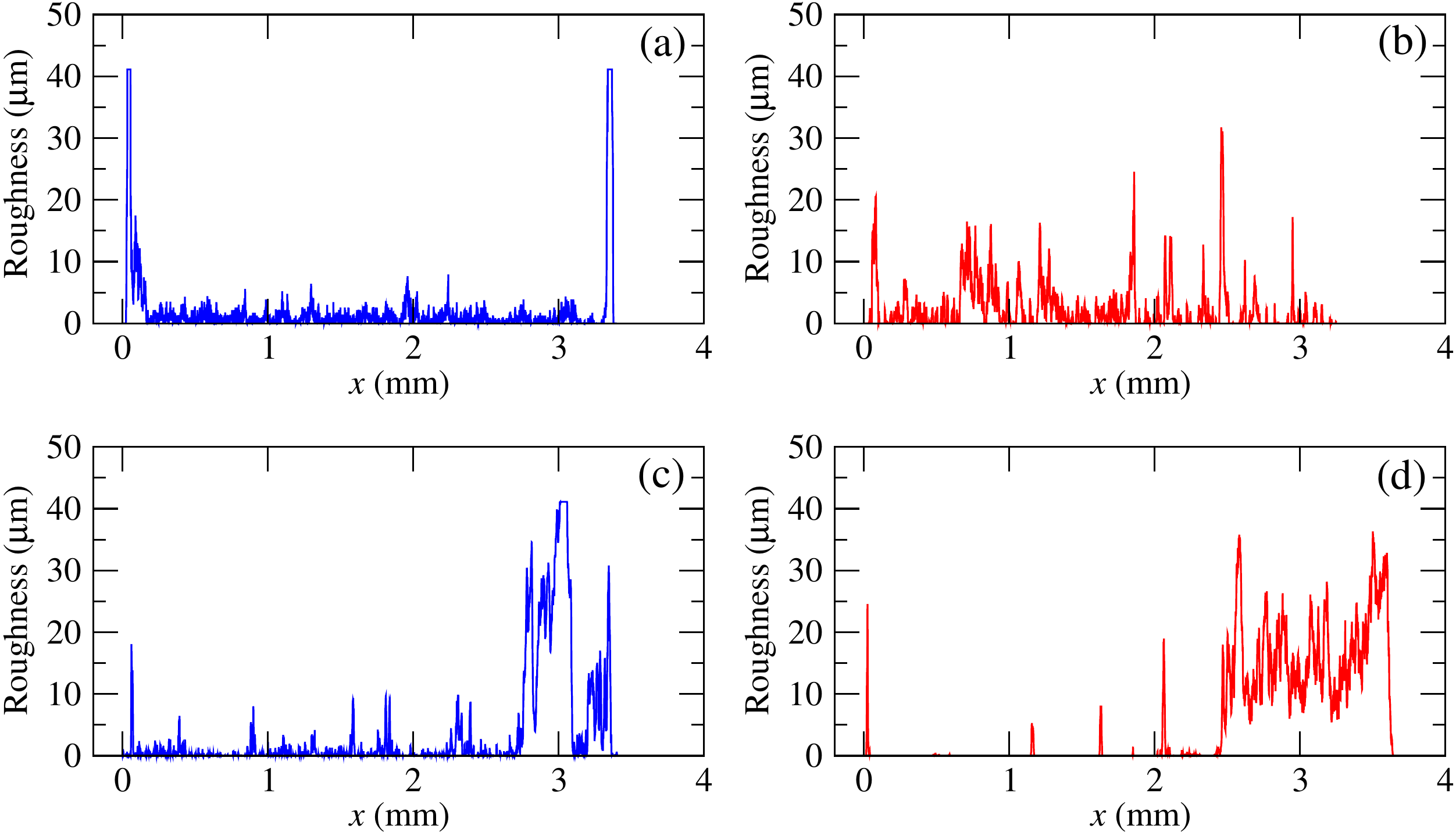}
\caption{Average roughness profile along a horizontal strip for the deposition pattern of ethanol droplets containing (a) 25 nm  Al$_2$O$_3$, (b) 25 nm Cu, (c) 75 nm Al$_2$O$_3$, and (d) 75 nm Cu nanoparticles.}
\label{fig:fig7}
\end{figure}

After a nanofluid droplet evaporates, the deposited nanoparticles form different patterns on the substrate depending on the evaporation and contact angle dynamics experienced by the droplet. The deposition patterns on the substrates after the droplets laden with Al$_2$O$_3$ and Cu nanoparticles of 25 nm and 75 nm are fully evaporated have been depicted in Figure \ref{fig:fig6}. The corresponding roughness profiles obtained with the help of a digital microscope are shown in Figure \ref{fig:fig7}. It is seen that for the droplet with Al$_2$O$_3$ (25 nm) nanoparticles, the deposits are concentrated mainly near the triple contact line. As the droplet evaporates, it experiences maximum evaporation at the triple line, which sets up a radial capillary flow from the droplet centre to the droplet periphery. The nanoparticles are displaced and brought to the triple contact line by this flow, where they pin the droplet. The Marangoni flow, which emerges as a consequence of the gradients in surface tension, also occurs in combination with the radial capillary flow. As a result, some of the nanoparticles are moved away from the triple contact line. As the evaporation continues, more particles are carried to the triple line by these radial flows, which gives a distinct pinned pattern, also known as the coffee ring \cite{deegan1997capillary,deegan2000contact,deegan2000pattern}. The droplets with Cu (25 nm), Al$_2$O$_3$ (75 nm), and Cu (75 nm) show stick-slip patterns. Even in these droplets, the radial and Marangoni flows prevail, but the stick-slip nature prevents most of the nanoparticles from depositing near the initial triple contact line. When the depinning occurs, the contact line gets displaced, where further deposits are found. Here the stick-slips pattern does not occur concentrically. Thus, we can observe a few locations around the initial contact line where the depositions are more, as in Figure \ref{fig:fig6}(c) (at the top-most portion of the droplet) and Figure \ref{fig:fig6}(d) (at two portions just above and below the heavily concentrated final deposition region). These correspond to the contact line portions that were shared by two or more droplet parts that experienced stick-slip. When comparing the droplet with Al$_2$O$_3$ (75 nm) and Cu (75 nm) nanoparticles, the droplet with Cu (25 nm) does not show an apparent stick-slip behaviour. The reason behind this behaviour is not known, and further investigation is required to address this issue.

The stick-slip behaviour could be explained with the help of a theoretical model developed by Shanahan \cite{shanahan1995simple}. It was suggested that the triple contact line has a potential energy barrier because of the mechanical (roughness) or chemical heterogeneity of the solid substrate. For an ideal substrate, Young's equation gives the equilibrium contact angle of the droplet. As the droplet evaporates, in order to minimise the surface free energy at any given moment, it would prefer to maintain the equilibrium contact angle, thus evaporating in a CCA mode. However, roughness and chemical heterogeneities provide an anchoring effect for the triple contact line. This anchoring could be associated with the potential energy barrier along the triple contact line. This anchoring prevents the droplet from attaining a state of minimum energy, and the excess energy is stored as the excess free energy in the droplet. As the droplet evaporates further, more excess energy is stored until it overcomes the potential energy barrier, where it slips to another equilibrium position. Thus, the droplet sticks until the excess energy in the droplet overcome the potential energy barrier and slips once it is equal to it, hence the occurrence of stick-slip dynamics. It can be said that during evaporation, the anchoring effect disturbs the capillary equilibrium, which in turn is responsible for the excess energy in the droplet. \citet{lin2016roughness} observed an increase in the potential energy barrier of polymeric substrates with an increase in roughness due to which the droplet is pinned to rough substrates.

Considering the deposition of nanoparticles along the triple contact line as a source of roughness, we can now compare the pinned and stick-slip behaviour of the droplets. For the Al$_2$O$_3$ (25 nm) droplet, the nanoparticles are deposited near the triple contact line, which increases the local roughness at that point. This increases the potential energy barrier along the triple contact line, which prevents the slipping of the droplet. The Al$_2$O$_3$ (25 nm) nanoparticles are lighter compared to other nanoparticles of Cu (25 nm), Al$_2$O$_3$ (75 nm), and Cu (75 nm) due to its smaller diameter and lower density as given in \ks{Table \ref{table:TS1}}. Due to this, the radial capillary forces can deposit more Al$_2$O$_3$ (25 nm) nanoparticles at the triple line. Thus the rate of increase of the potential energy barrier is faster than the rate of excess energy stored in the droplet, and thereby, the droplet cannot overcome the potential energy barrier and remains pinned throughout its lifetime. As for the case of Cu (25 nm), Al$_2$O$_3$ (75 nm), and Cu (75 nm), owing to the heaviness of these particles, the potential energy barrier could not be increased at a faster rate. Thus there are times when the excess energy in the droplet overcomes the barrier and slips to a new location with a contact angle less than or equal to the equilibrium contact angle. This process continues, which gives rise to the stick-slip pattern. 

The roughness profiles obtained by considering a small horizontal strip along the centre of the droplet are plotted and shown in Figure \ref{fig:fig7}. The dried deposition pattern of an ethanol droplet containing the nanoparticles was scanned using a digital microscope, and the corresponding images were processed using ImageJ software to obtain the roughness profile. It is seen that peaks from the roughness plots are in accordance with the deposition patterns in Figure \ref{fig:fig6}. The highest roughness at the triple contact line is given by the pinned droplet Al$_2$O$_3$ (25 nm), which relates to an increased deposition near the triple contact line compared with the other stick-slip droplets. The droplets with Cu (25 nm), Al$_2$O$_3$ (75 nm), and Cu (75 nm) have individual peaks which relate to the stick-slip patterns. In the droplets with Al$_2$O$_3$ (75 nm) and Cu (75 nm), after the complete evaporation of the droplet, the nanoparticles are deposited in a small region due to which they show areas of increased roughness ( rightmost portion of the drop). In the case of the droplet with Cu (25 nm), individual peaks along with a uniform deposition are present. The 3D images of the deposition pattern are shown in \ks{Figure S8}. 

\section{Conclusions}\label{sec:conclusions}
It is a dogma that increasing the thermal conductivity of a liquid by adding a small amount of nanoparticles increases the heat transfer rate \cite{eapen2007mean,cui2022enhanced,zhang2021anisotropic}. Thus, it has been used as a common strategy to enhance heat transfer in many applications, such as inkjet printing, fabrication of DNA microarrays, coating technology, spray and hotspot cooling and microfluidics. However, the universality of this result has been questioned by some researchers\cite{eapen2007mean}. Moreover, it is important to understand the mechanism underlying the improvement of heat transfer in nanofluid droplets. Thus, in the present study, the evaporation of sessile ethanol droplets laden with and without nanoparticles on a heated substrate is investigated using shadowgraphy and infrared (IR) imaging techniques considering Al$_2$O$_3$ and Cu nanoparticles of sizes 25 nm and 75 nm. The captured images are post-processed using \textsc{Matlab}$^{\circledR}$ and a machine learning technique. We found that the lifetime of a droplet is reduced by the addition of nanoparticles irrespective of particle type and size. However, the extent of loading has an insignificant effect on the evaporation time of the droplets. We observe that although the thermal conductivity of Al$_2$O$_3$ is an order of magnitude lower than that of Cu, droplets laden with 25 nm sized Al$_2$O$_3$ evaporate much faster than other droplets (droplets with 25 nm and 75 nm sized Cu nanoparticles, droplets with 75 nm sized Al$_2$O$_3$ nanoparticles). As the droplets with 25 nm Al$_2$O$_3$ nanoparticles exhibit pinned contact line dynamics while other droplets show stick-slip behaviour during the evaporation process, the counter-intuitive enhanced evaporation in the case of droplets with 25 nm Al$_2$O$_3$ can be attributed to the droplet contact line dynamics due to the presence of different nanoparticles. Additionally, the droplets with different nanoparticles exhibit distinct thermal patterns due to the difference in contact line behaviour, which alters the heat transfer inside the droplets. The temporal variations of the perimeter and free surface area, and the deposition patterns on the substrate for different loading conditions support our claim that the pinned contact line dynamics plays the dominant role in the enhanced heat transfer process rather than the increase in thermal conductivity of the nanoparticles. We believe that the light-weight 25 nm Al$_2$O$_3$ nanoparticles are more effectively transported to the triple contact line by radial thermo-capillary forces, resulting in more efficient pinned behaviour for these nanofluid droplets compared to the droplets with other type of nanofluid droplets considered in our study. Thus, the present study answers a fundamental prevalent question and provides a guideline for choosing the type and size of nanoparticles to increase the evaporation rate.  
\\

\noindent{\bf Credit authorship contribution statement} 

Hari Govindha and Pallavi Katre performed the experiments. All the authors contributed to the analysis of the results and to the preparation of manuscript. The project was coordinated by Kirti Chandra Sahu.\\

\noindent{\bf Declaration of Competing Interest} 

The authors declare that there is no conflict of interest.\\

\noindent{\bf Supporting Information:} 
\ks{Additional experimental details, Image processing steps and repeatability of experiments. Temporal evolution of the shape of the droplet and its wetted diameter for different conditions. Digital microscope images of the deposited nanoparticles.}

\noindent{\bf Acknowledgement:} {The financial support from Science \& Engineering Research Board, India through the grant number: CRG/2020/000507 is gratefully acknowledged.}


\providecommand{\latin}[1]{#1}
\makeatletter
\providecommand{\doi}
  {\begingroup\let\do\@makeother\dospecials
  \catcode`\{=1 \catcode`\}=2 \doi@aux}
\providecommand{\doi@aux}[1]{\endgroup\texttt{#1}}
\makeatother
\providecommand*\mcitethebibliography{\thebibliography}
\csname @ifundefined\endcsname{endmcitethebibliography}
  {\let\endmcitethebibliography\endthebibliography}{}


\end{document}